\documentclass[epj]{webofc}
\usepackage[utf8]{inputenc}
\usepackage[varg]{txfonts}   
\usepackage{booktabs}
\usepackage{xcolor}
\definecolor{darkred}{rgb}{0.4,0.0,0.0}
\definecolor{darkgreen}{rgb}{0.0,0.4,0.0}
\definecolor{darkblue}{rgb}{0.0,0.0,0.4}
\usepackage[bookmarks,linktocpage,colorlinks,
    linkcolor = darkred,
    urlcolor  = darkblue,
    citecolor = darkgreen]{hyperref}
\usepackage{bbm}
\usepackage{color}
\usepackage{tabularx, booktabs}
\newcolumntype{Y}{>{\centering\arraybackslash}X}
\newcommand{\beq}{\begin{equation}}
\newcommand{\eeq}{\end{equation}}
\newcommand{\bea}{\begin{eqnarray}}
\newcommand{\eea}{\end{eqnarray}}
\renewcommand{\d}{\delta}
\renewcommand{\l}{\lambda}

\renewcommand{\a}{\alpha}

\newcommand{\bx}{\mathbf{x}}
\newcommand{\Dt}{{\cal D}}
\newcommand{\on}{\frac{1}{9}}

\newcommand{\vx}{{\vec{x}}}
\newcommand{\vy}{{\vec{y}}}

\newcommand{\vk}{{\vec{k}}}

\newcommand{\tK}{\widetilde{K}}

\newcommand{\tr}{\mbox{Tr}}

\newcommand{\m}{\mu}

\newcommand{\D}{\Delta}

\newcommand{\oh}{\frac{1}{2}}

\newcommand{\dg}{\dagger}
\newcommand{\non}{\nonumber}

\newcommand{\rf}[1]{(\ref{#1})}
\newcommand{\ra}{\rightarrow}
\newcommand{\pa}{\partial}

\usepackage{subfigure}
\wocname{EPJ Web of Conferences}
\woctitle{Lattice2017}
\begin{document}
%
\selectlanguage{english}
\title{%
Preliminary QCD phase transition line for 695 MeV dynamical staggered fermions from effective Polyakov line actions}
\author{%
\firstname{Roman}
\lastname{H\"ollwieser}\inst{1}\inst{2}\fnsep\thanks{Speaker at Lattice
2017 in Granada, \email{hroman@kph.tuwien.ac.at}, supported by DFG under SFB/TRR55.}
\and
\firstname{Jeff}  \lastname{Greensite}\inst{3}\fnsep
}
\institute{%
Nuclear Physics Department, Vienna University of Technology, Operngasse 9, 1040 Vienna, Austria
\and
Theoretical Particle Physics, Bergische Universit\"at Wuppertal, Gau{\ss}str. 20, 42119 Wuppertal, Germany
\and
Physics and Astronomy Department, San Francisco State University, San Francisco, CA~94132, USA
}

\abstract{%
We present a phase diagram for SU(3) lattice gauge theory with 695 MeV dynamical staggered fermions, in the plane of temperature and chemical potential, obtained from effective Polyakov line actions. The derivation is via the method of relative weights, and the effective theories are solved at finite chemical potential by mean field theory.
}
\maketitle
\section{Introduction}\label{intro}

    One of the most active areas in strong interaction physics concerns the behavior of QCD in extreme conditions, {\it i.e.}, high temperature and/or high baryon density.   At high temperatures we enter the realm of the quark gluon plasma, whose properties have been or will be probed by experiments at RHIC, the LHC, and the FAIR facility (now under construction). Not much is known for sure about hadronic matter at high baryon density.  QCD is believed to have a complex phase structure in the temperature-density plane, and a number of exotic phases (quarkyonic, glasma, color-flavor locked superconductor...) have been suggested .  One would especially like to know the position of the critical end point of the confinement-deconfinement transition.  Many talks on the subject of QCD in extreme environments begin with a sketch of the possible phase diagram, but such sketches are, so far, all conjecture.  Nobody knows whether these exotic phases really exist, or exactly where they are located in the temperature/density plane.  So the first order of business, for a theorist, is to nail down the phase diagram.
By far the most important tool in the investigation of non-perturbative QCD is the method of importance sampling in lattice Monte Carlo simulations. But when one attempts to apply this tool to study QCD at high baryon density, a serious obstacle $-$ the ``sign problem'' $-$ is encountered. Different strategies were explored, and the reviews at the yearly lattice conferences \cite{Ding:2017giu,Borsanyi:2015axp,Sexty:2014dxa,Gattringer:2014nxa,Aarts:2013lcm,Levkova:2012jd,Wolff:2010zu,deForcrand:2010ys,Chandrasekharan:2008gp} summarize the progress. Finite densities are introduced in statistical systems via the introduction of a chemical potential, but when this standard method is applied in QCD, the fermion determinant becomes complex and cannot be interpreted as a probability measure.  Then standard importance sampling, {\it e.g.}, via the hybrid Monte Carlo method, breaks down completely, and some other method or methods must be devised to handle the problem of complex actions. The sign problem crops up not only in QCD at high baryon density, but also in various condensed matter systems involving highly correlated fermions.  So apart from the general conceptual challenge, it would be rewarding in many areas of physics if we could figure out how to apply numerical methods to systems with complex actions. There are many promising avenues, of course, in particular the complex Langevin equation~\cite{Sexty:2013ica}, Lefshetz thimbles~\cite{Cristoforetti:2013wha}, and the histogram approach~\cite{Ejiri:2013lia}. However, no method so far is entirely free from objections, {\it e.g.}, in the case of the complex Langevin equation, the problem is that the fermionic part of the action is non-holomorphic, due to the fact that the logarithm of the fermion determinant has a branch cut.  It has been shown by Mollgaard and Splittorff~\cite{Mollgaard:2013qra}, that this can lead complex Langevin evolution to incorrect results. As for the Lefshetz thimble approach, the rigorous statement is that all thimbles must be summed over, but the proposed application to  QCD is to simulate quantum fluctuations around a single thimble.  It is not yet clear that this is the right thing to do, and in simple examples the restriction to a single thimble gives the wrong answer.  Finally, the histogram approach was studied critically by Joyce Myers, Kim Splittorff, and Jeff Greensite~\cite{Greensite:2013gya,Greensite:2013ska}. What was found is that the fundamental assumption of the histogram approach, namely, that the phase of the fermion determinant has a Gaussian distribution, is probably not right, and that even tiny deviations from a Gaussian distribution, on the order of inverse powers of the lattice volume, are sufficient to invalidate conclusions based on a Gaussian distribution alone.

Our approach to the sign problem in QCD is to map QCD with a chemical potential into a simpler effective theory, namely, the effective Polyakov line action (henceforth ``PLA''), via the relative weights method and then deal with the sign problem via mean field theory, which is a surprisingly accurate method for solving effective actions of this kind~\cite{Greensite:2013gya}. The general idea was pioneered in \cite{Fromm:2011qi}, and the derivation of the PLA from the underlying LGT has been pursued by various methods, e.g.\ \cite{Gattringer:2011gq,Bergner:2015rza,bergner:2013qaa,scior:2014zga,Scior:2016fso,Bahrampour:2016qgw,Wozar:2007tz}.  The relative weights method \cite{Greensite:2014isa,Greensite:2013bya} is a simple numerical technique for finding the derivative of the PLA in any direction in the space of Polyakov line holonomies. Given some ansatz for the PLA, depending on some set of parameters, we can use the relative weights method to determine those parameters.  Then, given the PLA at some fixed temperature $T$, we can apply a mean field method to search for phase transitions at finite chemical potential $\m$.  The phase diagram of the effective theory will mirror the phase diagram of the underlying gauge theory. The method was successfully tested in SU(2) and SU(3) pure gauge and gauge-Higgs theories \cite{Greensite:2014isa,Greensite:2013bya,Greensite:2013yd} and first results with dynamical fermions were presented in~\cite{Hollwieser:2015nna,Hollwieser:2016hne,Greensite:2016fha,Hollwieser:2016yjz}. Here we follow up on this work and apply the relative weights method to SU(3) gauge theory with staggered fermions of mass 695 MeV at a set of temperatures in the range $129 \le T \le 260$, to obtain an effective Polyakov line action at each temperature.  We then apply a mean-field method to search for phase transitions in the effective theory at finite densities. We obtain a tentative transition line in the $\m-T$ plane that has an endpoint at high temperatures, as expected, and a second, unexpected endpoint at a lower temperature. It remains to be seen, whether a transition line reappears at still lower temperatures, or whether the second transition point disappears for lighter quarks, or whether this second transition point is instead indicative of some deficiency in our ansatz for the PLA. See also~\cite{Greensite:2017qfl} for more details.

\section{Polyakov Line Action and Relative Weights}
 
\begin{figure}[h]
	\centering
	a)\includegraphics[width=.477\linewidth]{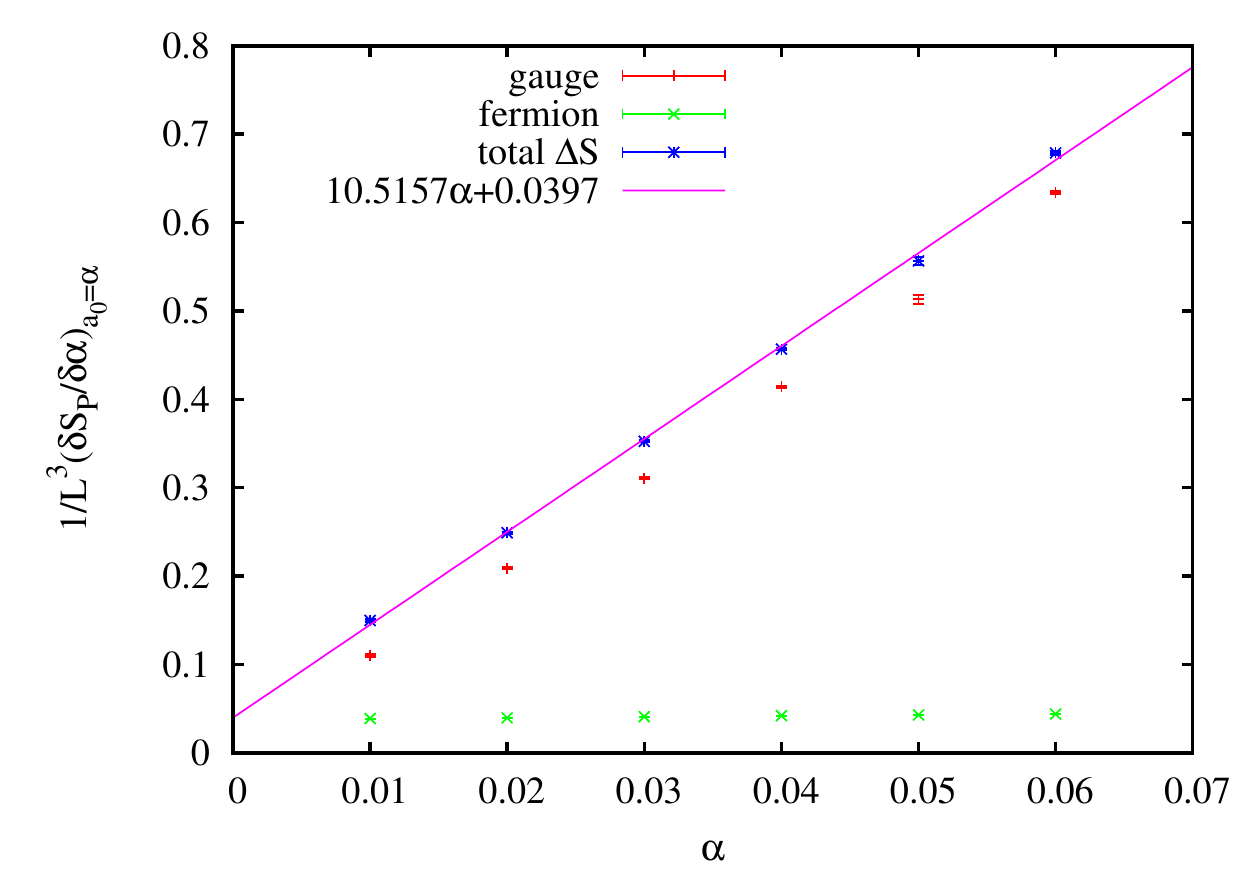}
	b)\includegraphics[width=.477\linewidth]{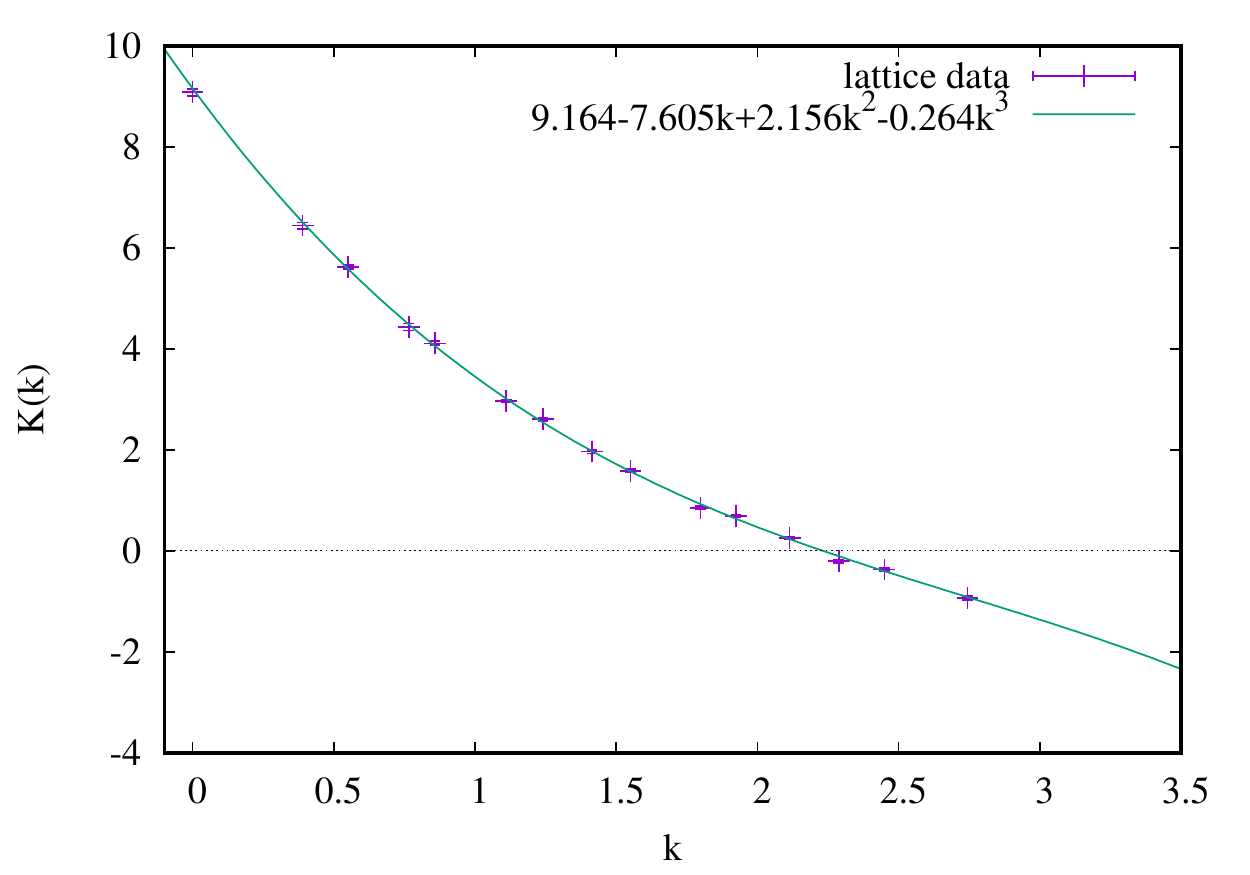}
\caption{a) Derivatives of $S_P$ with respect to momentum modes $a_\vk$, evaluated at $a_\vk=\a$ for the zero mode $k=0$: data points fit by a straight line of slope $2\tK(0)\alpha$, b) The kernel $K(k)$ in momentum space fit by a cubic function.}
\label{fig:dSKk}
\end{figure}

\begin{figure}
\centering
	a)\includegraphics[width=.477\linewidth]{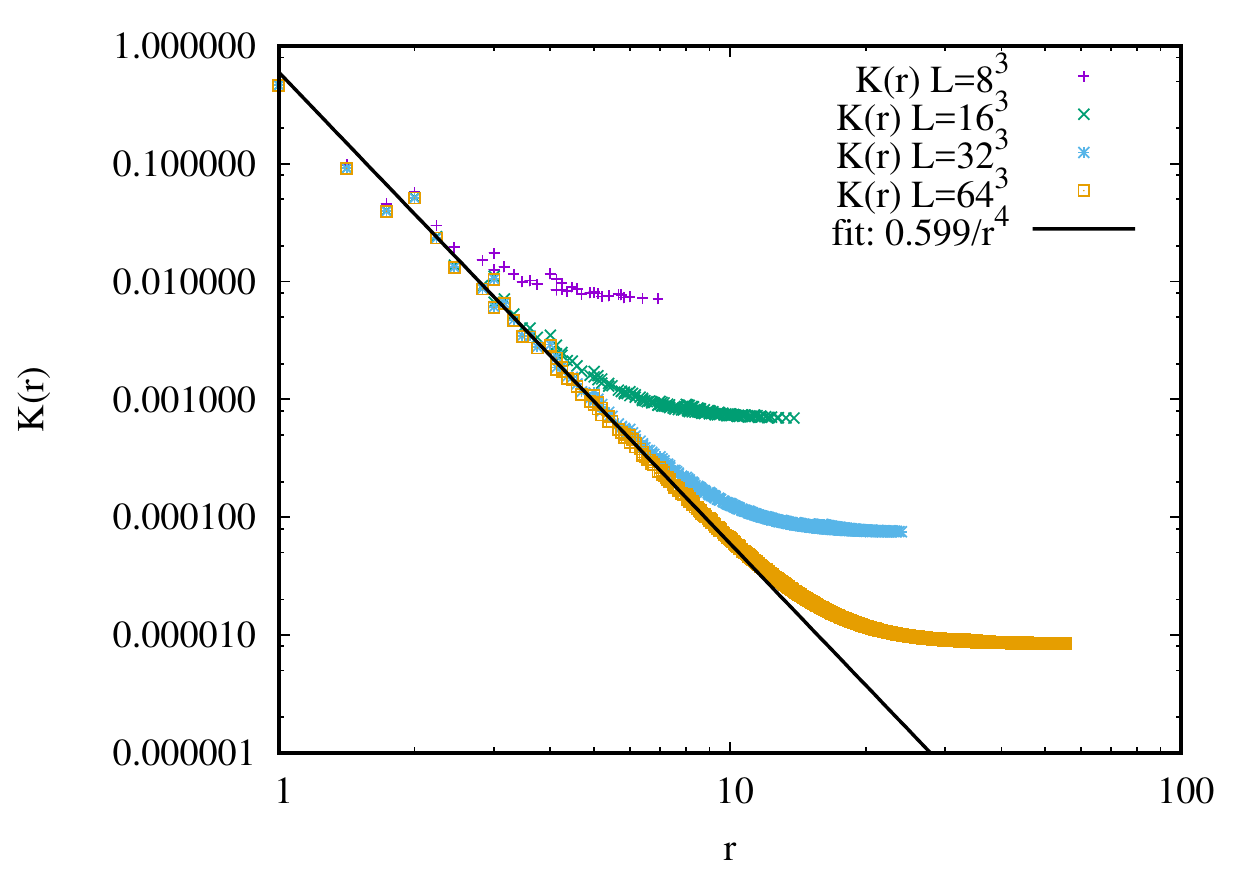}
	b)\includegraphics[width=.477\linewidth]{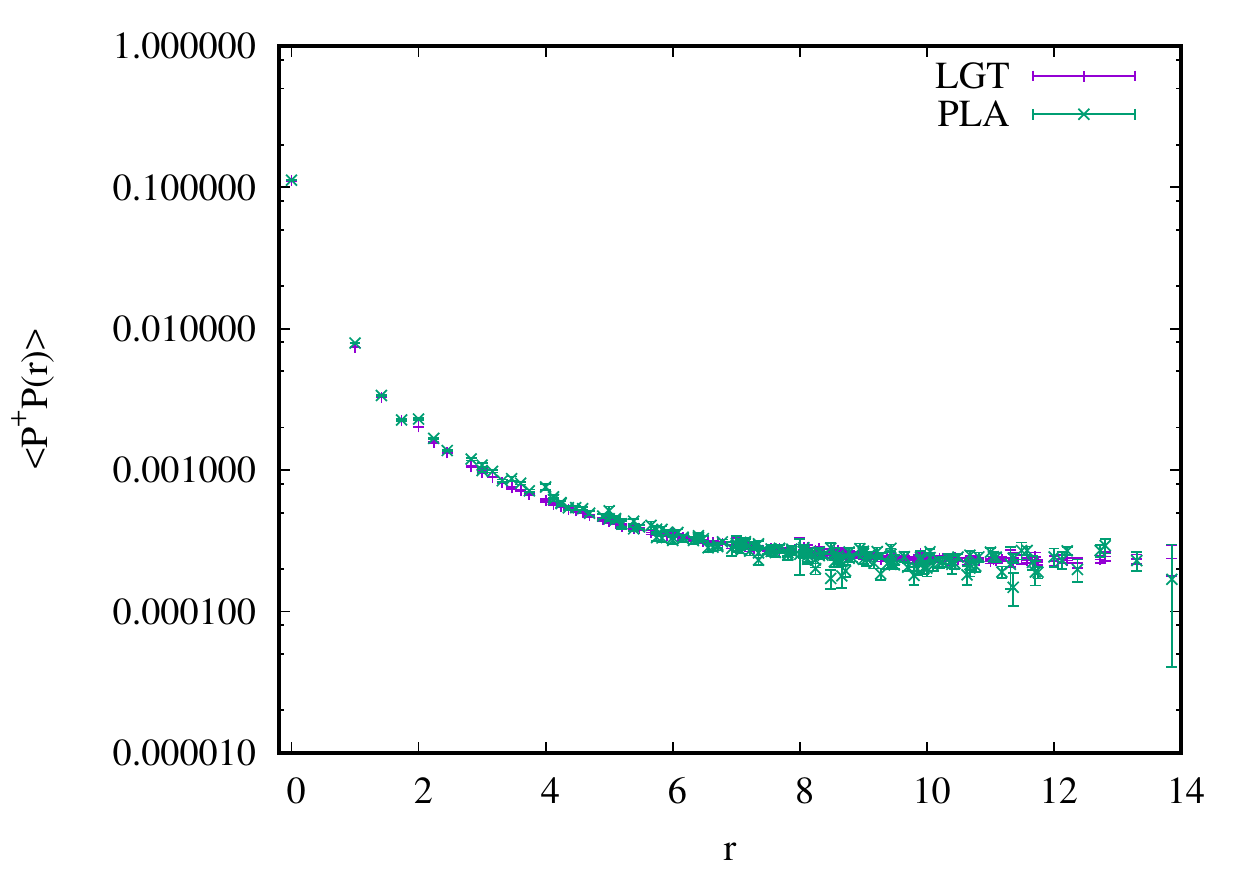}
\caption{a) The kernel $K(r)=K(\vx-\vy)$ of the PLA for various lattice sizes and an infinite volume fit. b) Polyakov line correlators from lattice and PLA simulations.}
\label{fig:krpla}
\end{figure}

The effective Polyakov line action $S_P$ is the theory obtained by integrating out all degrees of freedom of the lattice gauge theory, under the constraint that the Polyakov line holonomies are held fixed.  It is convenient to implement this constraint in temporal gauge (${U_0(\bx,t\ne 0)=\mathbbm{1}}$), so that
\bea
\exp\Bigl[S_P[U_{\vx},U^\dg_{\vx}]\Bigl] =  \int  DU_0(\vx,0) DU_k  D\phi \left\{\prod_{\vx} \d[U_{\vx}-U_0(\vx,0)]  \right\} e^{S_L} \ ,
\label{S_P}
\eea
where $\phi$ denotes any matter fields, scalar or fermionic, coupled to the gauge field, and $S_L$ is the SU(3) lattice action (note that we adopt a sign convention for the Euclidean action such that the Boltzmann weight is proportional to $\exp[+S]$).  
To all orders in a strong-coupling/hopping parameter expansion, the relationship between the PLA at zero chemical potential 
$\m=0$, and the PLA corresponding to a lattice gauge theory at finite chemical potential, is given by
\bea
     S_P^\m[U_\vx,U^\dg_\vx] =  S_P^{\m=0}[e^{N_t \m} U_\vx,e^{-N_t \m}U^\dg_\vx]   \ .
\label{convert}
\eea
So the immediate problem is to determine the PLA at $\m=0$.
Let us define the Polyakov line in an SU($N$) theory to refer to the trace of the Polyakov line holonomy
$P_\vx \equiv  {1 \over N} \tr[U_\vx]$.
Relative weights  enables us to compute the derivative of the effective action $S_P$ along any path
\beq
            \left( {d S_P \over d \l}\right)_{\l=\l_0} \approx {\D S_P \over \D \l} 
\eeq
at any point $\{U_\vx(\l_0)\}$ in the configuration space of all $U_\vx$ on the lattice volume, parametrized by $\l$. We compute the derivatives of the effective action, by the relative weights method, with respect to the
Fourier (``momentum'') components $a_\vk$ of the Polyakov line configurations $P_\vx = \sum_\vk  a_\vk e^{i \vk \cdot \vx}$. Components of the wavevector $k_i = 2\pi m_i/L$ are specified by a triplet of integer mode numbers 
$(m_1,m_2,m_3)$, and in this work  we have used triplets
\bea 
&& (000),(100),(110),(200),(210),(300),(311),(400),(322),(430),(333),
\non\\
&& (433),(443),(444),(554),(6,5,4),(6,6,6),(7,6,5),(8,6,7),(8,8,8)
\label{eq:ks}
\eea
with spatial lattice extension $L=16$.
The procedure is to run a standard Monte Carlo simulation, generate a configuration of Polyakov line holonomies  $U_\vx$, and compute the Polyakov lines $P_\vx=\tr U_\vx$.  We then set the momentum mode $a_\vk=0$ in this configuration to zero, resulting in the modified configuration $\widetilde{P}_\vx =  P_\vx - \left({1\over L^3} \sum_\vy P_\vy e^{-i\vk \cdot \vy}\right) e^{i\vk \cdot \vx}$, and define
\bea
            P''_\vx = \Bigl(\a - \oh \D \a \Bigr) e^{i\vk \cdot \vx} + f \widetilde{P}_x \ \quad,\quad
            P'_\vx = \Bigl(\a + \oh \D \a \Bigr) e^{i\vk \cdot \vx} + f \widetilde{P}_x \ ,
\eea
where $f$ is a constant close to one ($f=1$ is only possible in the large volume, $\a \ra 0$ limit). 
From the holonomy configurations $U''_x, U'_x$ we compute derivatives of $S_P$ with respect to the real part $a^R_\vk$ of the Fourier components $a_\vk$ by defining $\D S_P = S_P[U'_\vx] - S_P[U''_\vx]$, hence we have from \rf{S_P},
\bea
e^{\D S_P} =  {\int  DU_k  D\phi ~  e^{S'_L} \over \int  DU_k  D\phi ~  e^{S''_L} }
= {\int  DU_k  D\phi ~  \exp[S'_L-S''_L] e^{S''_L} \over \int  DU_k  D\phi ~  e^{S''_L} }
= \Bigl\langle  \exp[S'_L-S''_L] \Bigr\rangle'' \ .
\eea
The expectation value is straightforward to compute numerically, by fixing the Polyakov holonomies and calculating the action differences, and from the logarithm we determine $\D S_P$.
Our proposal is to fit the relative weights data to an ansatz for $S_P$ based on the massive quark effective 
action~\cite{Fromm:2011qi,Bender:1992gn,Blum:1995cb,Engels:1999tz,DePietri:2007ak} 
\bea
S_P[U_\vx] &=& \sum_{\vx,\vy} P_\vx K(\vx-\vy) P_\vy
+p \sum_\vx \bigg\{ \log(1+he^{\mu/T}\tr U_\vx+h^2e^{2\mu/T}\tr U_\vx^\dagger+h^3e^{3\mu/T}) \non \\
& & \qquad +\log(1+he^{-\mu/T}\tr U_\vx+h^2e^{-2\mu/T}\tr U_\vx^\dagger+h^3e^{-3\mu/T}) \bigg\} \ ,
\label{eq:SP}
\eea
where both the kernel $K(\vx-\vy)$ and the parameter $h$ are to be determined by the relative weights method.  The full action is surely more complicated than this ansatz; the assumption is that these terms in the action are dominant, and the effect of a lighter quark mass is mainly absorbed into the parameter $h$ and kernel $K(\vx-\vy)$. We then have the derivative of the action with respect to momentum modes $a_\vk$ of the Polyakov lines
\bea
 {1\over L^3}\left( {\pa S_P \over \pa a^R_\vk} \right)_{a_\vk=\a} = 2 \tK(\vk) \a
+ {p\over L^3} \sum_\vx \left( 3h e^{i \vk \cdot \vx} + 3h^2  e^{-i \vk \cdot \vx} + \mbox{c.c} \right) \ .
\eea
The left-hand side is computed via relative weights at a variety of $\a=0.01,\ldots,0.06$, and plotting those results vs.\ $\a$, $K(\vk)$ is determined from the slope, while $h$ is given by the intercept at $\a=0$ from the zero mode $k=0$, since the second term basically vanishes for $k\ne0$, see Figure~\ref{fig:dSKk}. We fine-tune $h$ by requiring that $\langle P \rangle$ agrees in the PLA with the underlying LGT. From the kernel $K(\vk)$ in momentum space we derive the kernel $K(\vx-\vy)$ via inverse Fourier transformation and find that it behaves like $K(r)=c/r^4$, see Figure~\ref{fig:krpla}a. At finite lattice spacing it deviates from the simple ansatz for large $r$, hence we introduce a cutoff $R_{cut}$ and set the kernel to zero for $r>R_{cut}$ in our PLA simulations. A first consistency check of our PLA simulations is that we reproduce the correct Polyakov line correlator of the underlying lattice gauge theory, as shown in Figure~\ref{fig:krpla}b for $\beta=5.7$.

\section{Finite Density and Mean Field Theory}

For $\mu\ne0$ the effective PLA has still a sign problem, which we solve via mean field theory as discussed in \cite{Greensite:2012xv,Greensite:2014cxa}. The treatment of $SU(3)$ spin models at finite $\mu$ is just a minor variation of standard mean field theory at zero chemical potential. The basic idea that each spin is effectively coupled to the average spin on the lattice, not just nearest neighbors, is favored by the effective PLA with the non-local kernel $K(\vx-\vy)$. We introduce two magnetizations $u$ and $v$ for $\tr U_\vx$ and $\tr U_\vx^\dagger$ which are determined by minimizing the free energy. Therefore we rewrite the kernel part of the action (\ref{eq:SP}) to separate local and non-local terms
\beq
S_0 = J_0 \sum_\vx (v \tr U_\vx + u \tr U_\vx^\dg) - uvJ_0V + a_0 \sum_{\vx} \tr[U_\vx] \tr[U_\vx^\dg]+E_0 \ ,
\eeq
where $V=L^3$ is the lattice volume, and we have defined
\beq
     E_0 = \sum_{(\vx \vy)} (\tr U_x-u)(\tr U_\vy^\dg - v) \on K(\vx-\vy) \ , \quad
     J_0 =  {1\over 9} \sum_{\vx \ne 0} K(\vx) \ , \quad a_0 = {1\over 9} K(0) \ .
\eeq
$u$ and $v$ are chosen such that $E_0$ can be treated as a perturbation, in particular, $\langle E_0 \rangle = 0$ when $u = \langle \tr U_x \rangle$, $v = \langle \tr U^\dg_x \rangle$, and these conditions turn out to be equivalent to the stationarity of the mean field free energy.  The mean field approximation is obtained, at leading order, by dropping $E_0$, in which case the partition function factorizes, and can be solved analytically as a function of $u$ and $v$.  After some manipulations (cf.\ \cite{Greensite:2012xv,Greensite:2014cxa}), one finds the mean field approximations $u, v$ to $\langle \tr U_x \rangle$ and $\langle \tr U^\dg_x \rangle$ respectively, by solving the pair of equations
\beq
u - {1\over G}{\pa G \over \pa A}=0  ~~~~~\text{and}~~~~~ v - {1\over G}{\pa G \over \pa B}=0 \ ,
\label{hq-conditions}
\eeq
where $A=J_0 v, ~ B=J_0 u$.  The expression $G(A,B)$ is given by
\beq
G(A,B) =  \Dt\left(\m,{\pa \over \pa A},{\pa \over \pa B} \right) 
 \sum_{s=-\infty}^{\infty}   \det\Bigl[D^{-s}_{ij} I_0[2\sqrt{A B}] \Big]  \ ,
\label{G}
\eeq
where $D^{-s}_{ij}$ is the $i,j$-th component of a matrix of differential operators
\beq
D^s_{ij} = \left\{ \begin{array}{cl}
                         D_{i,j+s} & s \ge 0 \cr
                         D_{i+|s|,j} & s < 0 \end{array} \right. \ , \quad
D_{ij} = \left\{ \begin{array}{cl}
                         \left({\pa \over \pa B} \right)^{i-j} & i \ge j \cr 
                        \left({\pa \over \pa A} \right)^{j-i} & i < j \end{array} \right. \ .
\eeq
The mean field free energy density $f_{mf}$ and fermion number density $n$ are
\beq
           {f_{mf} \over T} = J_0 u v - \log G(A,B) ~~~~~\text{and}~~~~~ n = {1\over G} {\pa G \over \pa \m} \ .
\eeq

   The stationarity conditions \rf{hq-conditions} may have more than one solution, and here it is important to take account of the existence of very long lived metastable states in the PLA.  The state at $\m=0$ which corresponds to the LGT is the one obtained by initializing at $P_\vx=0$, and in the mean field analysis this is actually not the state of lowest free energy (its stability in a Monte Carlo simulation is no doubt related to the highly non-local couplings in the PLA).  By analogy, at finite $\m$ we look for solutions of \rf{hq-conditions} by starting the search at $u=v=0$, regardless of whether another solution exists at a slightly lower $f_{mf}$. A first check of the mean field approach is that we reproduce the correct expectation value of the Polyakov loop from LGT at $\mu=0$.

\section{The QCD Phase Transition Line}

\begin{figure}
\centering
	a)\includegraphics[width=.47\linewidth]{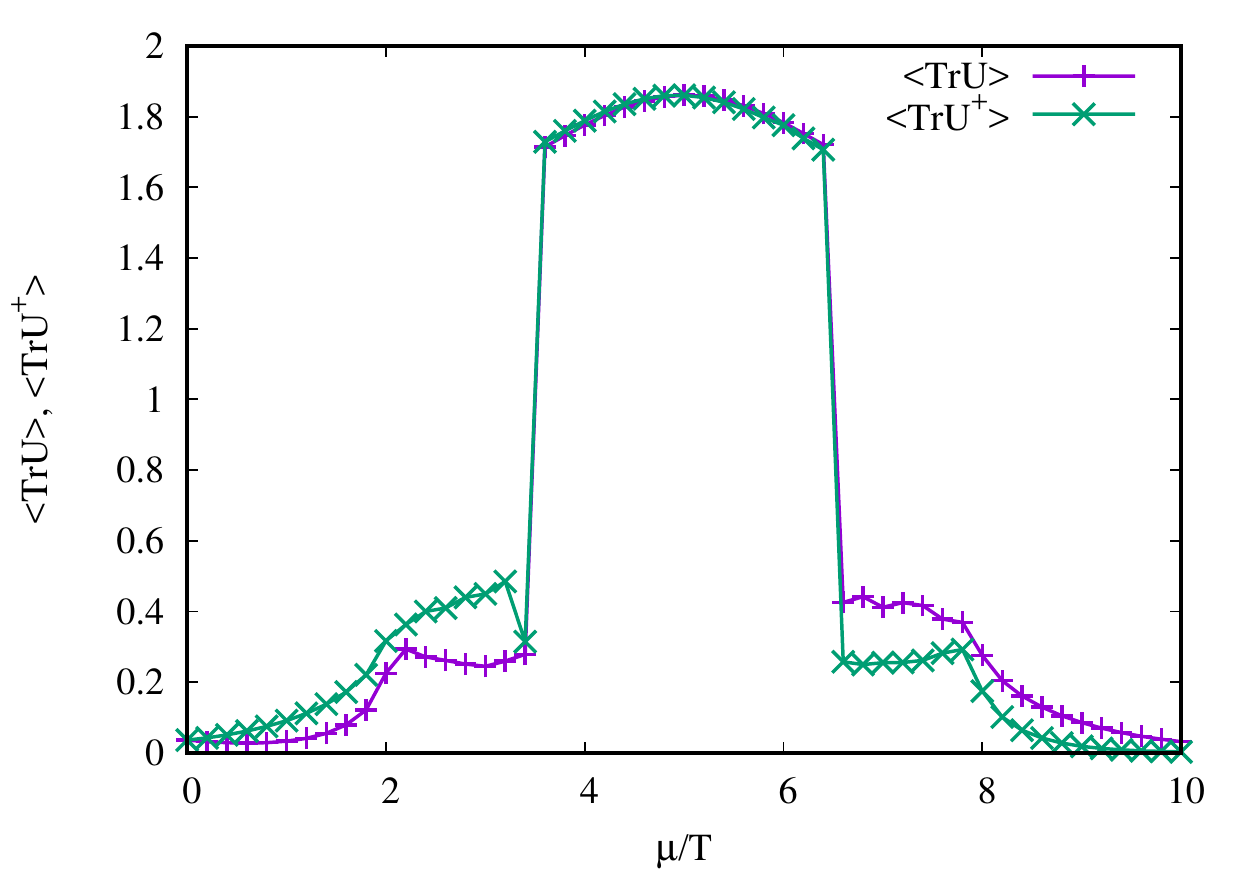}
	b)\includegraphics[width=.47\linewidth]{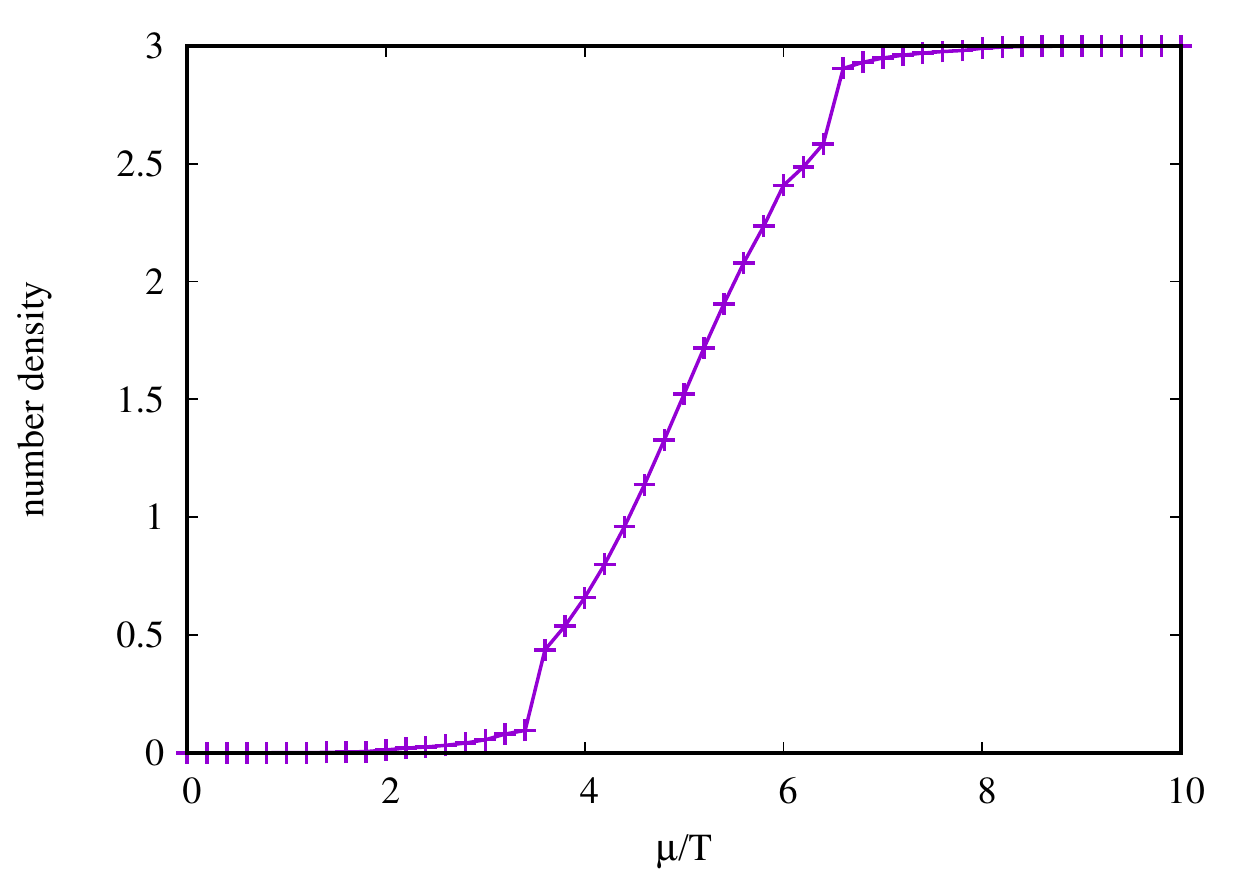}
\caption{a) Polyakov lines and b) number density from mean field analysis of the PLA derived from LGT at $\beta=5.7$ Wilson gauge coupling and dynamical staggered fermions. b) Preliminary phase diagram for $m_q=695$MeV. }
\label{fig:uvnd}
\end{figure}

We derive effective Polyakov line actions from lattice simulations of SU(3) Wilson gauge action and dynamical staggered fermions on $16^3 \times 6$ lattices, for a variety of temperatures and lattice masses corresponding to a physical quark mass $m_q=695$MeV, as summarized in Table~\ref{tab:par}. To set the scale we use the lattice spacing $a$ from the Necco-Sommer formula \cite{Necco:2001xg}.
We keep the physical mass and the extension $N_t=6$ in the time direction fixed, and vary the temperature by varying the lattice spacing, i.e.\ by varying $\beta$. Polyakov loop correlators in the LGT and the PLA at $\mu=0$ are in very good agreement in all cases as well as the Polyakov line expectation values after mean field treatment. Turning on the chemical potential $\m$, $\langle \tr U_\vx \rangle$ and $\langle \tr U^\dg_\vx \rangle$ are calculated by the mean field method outlined above, with a sample of our results displayed in Fig.\ \ref{fig:uvnd}.  A discontinuity is the sign of a transition at finite density, and conversely the absence of any discontinuity indicates the absence of any transition.  When a transition occurs at some value of chemical potential $\m_1$, then there is a second transition at some $\m_2 > \m_1$.   However, while the first transition occurs at some relatively low density (in lattice units) on the order of $n \approx 0.1$, the second transition always occurs at a density close to the saturation value, which for staggered fermions is $n=3$. Since the saturation value is a lattice artifact, we do not attach much physical significance to the second transition.

\begin{table}
\small
\begin{tabularx}{1.\textwidth}{|c *{10}{|Y}|}
\hline
$\beta$&$T$[MeV]&$a$[fm]&$ma$&P (Lgt)&P (mfd)&$R_{cut}$&$h$&$\mu_1/T$&$\mu_2/T$\\
\hline
5.55 & 129 & 0.248 & 0.875 & 0.00102 & 0.00101 & 5.5 & 0.0018 & - & - \\
5.60 & 147 & 0.217 & 0.767 & 0.00135 & 0.00133 & 5.5 & 0.0014 & - & - \\
5.63 & 158 & 0.201 & 0.711 & 0.00188 & 0.00189 & 5.5 & 0.0017 & 4.5 & 8.5 \\
5.65 & 167 & 0.192 & 0.677 & 0.00254 & 0.00249 & 5.5 & 0.0019 & 4.2 & 8.3 \\
5.70 & 188 & 0.170 & 0.601 & 0.01198 & 0.01195 & 5.0 & 0.0069 & 3.5 & 6.5 \\
5.73 & 201 & 0.159 & 0.561 & 0.05734 & 0.05731 & 5.0 & 0.0220 & 2.9 & 5.5 \\
5.75 & 211 & 0.152 & 0.536 & 0.07235 & 0.07110 & 5.0 & 0.0272 & 2.1 & 5.3 \\
5.77 & 220 & 0.145 & 0.513 & 0.08354 & 0.08347 & 5.0 & 0.0387 & 1.1 & 5.1 \\
5.775 & 222 & 0.144 & 0.508 & 0.08522 & 0.08523 & 4.0 & 0.0565 & - & - \\
5.78 & 225 & 0.142 & 0.502 & 0.08703 & 0.08679 & 4.0 & 0.0610 & - & - \\
5.80 & 235 & 0.136 & 0.482 & 0.09332 & 0.09437 & 4.0 & 0.0710 & - & - \\
5.85 & 260 & 0.123 & 0.435 & 0.10992 & 0.01115 & 4.0 & 0.0900 & - & - \\
\hline
\end{tabularx}
\caption{Simulation parameters for Wilson gauge action and dynamical
staggered fermions with $m_q=695$MeV and corresponding effective Polyakov line
actions on $16^3$x$6$ lattices together with mean field results.}\label{tab:par}
\end{table}

\begin{figure}
\centering
	a)\includegraphics[width=.47\linewidth]{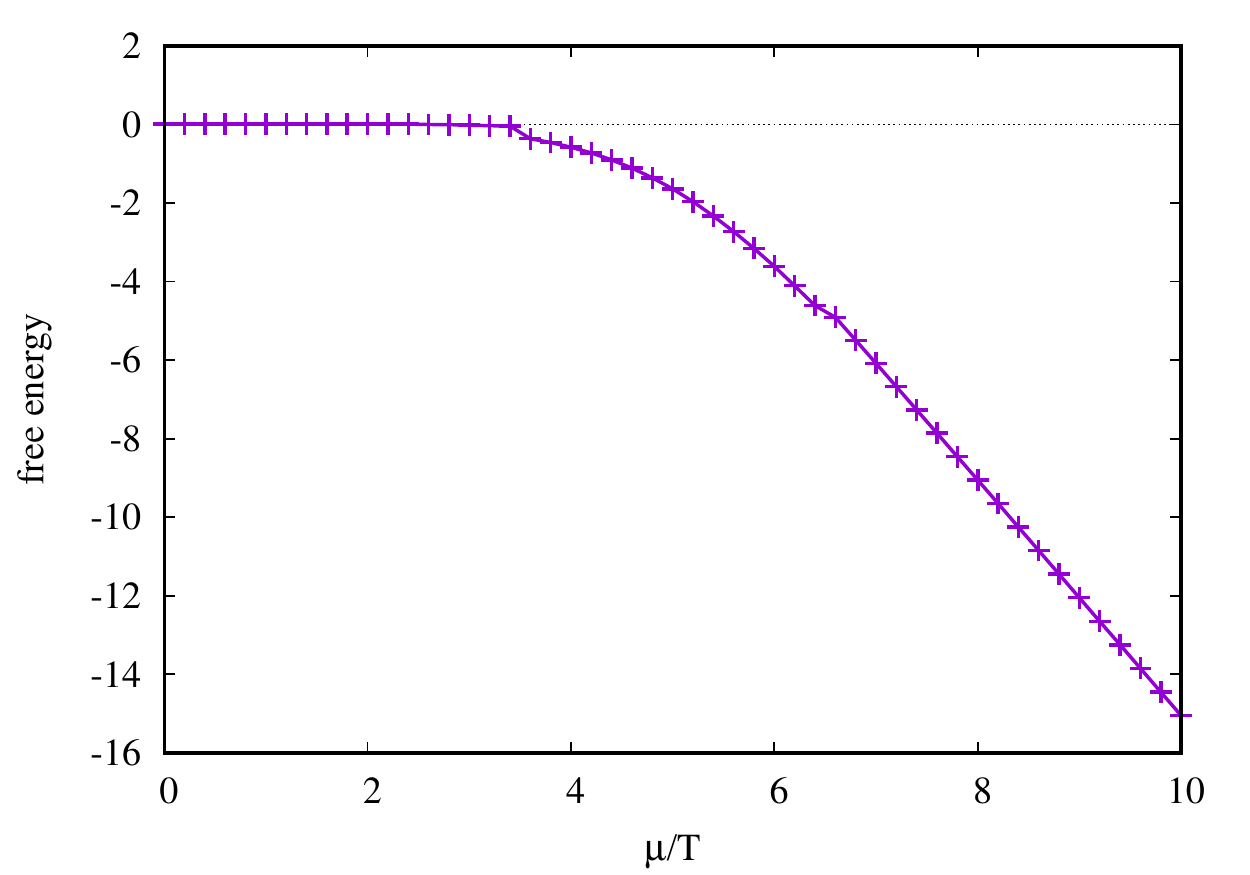}
	b)\includegraphics[width=.47\linewidth]{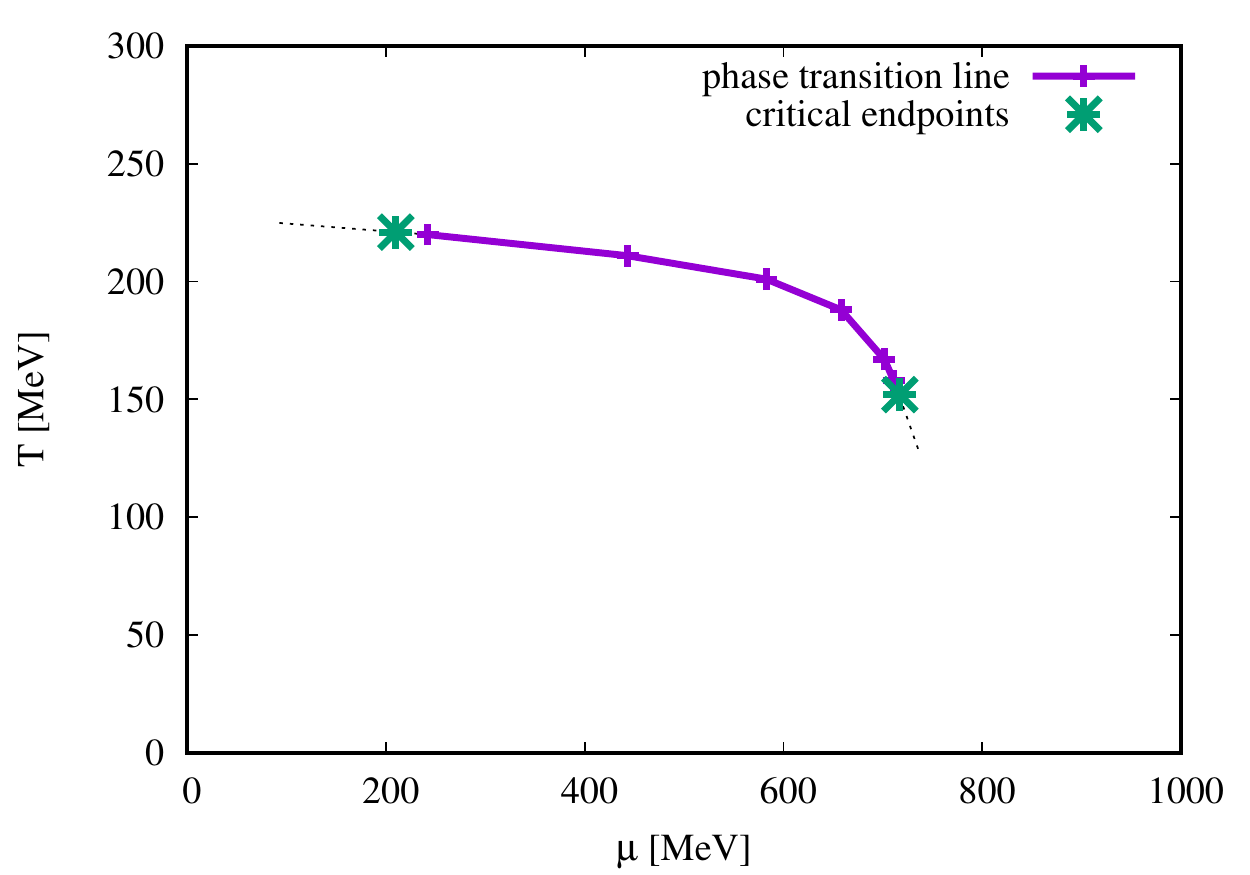}
\caption{a) Free energy from mean field analysis of the PLA derived from LGT at $\beta=5.7$ Wilson gauge coupling and dynamical staggered fermions. b) Preliminary phase diagram for $m_q=695$MeV. }
\label{fig:frpt}
\end{figure}

   From the $\m_1$ vs.\ $T$ data shown in the table, we plot a transition line in the $\m-T$ plane for staggered, unrooted quarks of mass 695 MeV, Fig.\ \ref{fig:frpt}b.  This figure is the main result of our paper, and holds for the temperature range $129 \le T \le 260$ MeV that we have
investigated.  We see that the phase transition line exists to an upper temperature of $T \approx 220$ MeV, where there is a critical
endpoint.  The fact that there is a critical endpoint at high temperature was expected.  What was unexpected is the existence of a second critical endpoint at a lower temperature of $T \approx 158$ MeV.  We cannot rule out the possibility that a high density transition reappears at some temperature lower than the lowest temperature (129 MeV) that we have considered.   Or perhaps the second critical endpoint goes away for light quarks. These possibilities we reserve for later investigation.   
    
Finally, we take a look at an interesting observable introduced in~\cite{Caselle:2017xrw}, the ratio $\xi/\xi_2$ of correlation length and second momentum, which tests for the presence, in an effective Polyakov line action, of a spectrum of excitations contributing to two-point correlators. Our values 
are comparable to some of the results for pure SU(2) lattice gauge theory quoted in ref.\ \cite{Caselle:2017xrw}, for more details and a thorough discussion see~\cite{Greensite:2017qfl}. 
    
\section{Conclusions}

    We have found a first-order phase transition line for SU(3) gauge theory with dynamical unrooted staggered fermions of mass 695 MeV, by the method of relative weights combined with mean field theory, in the plane of chemical potential $\m$ and temperature $T$.   This line has two critical endpoints, at $T=158$ MeV and $T=220$ MeV. It would also be interesting to see what happens to the second critical endpoint in a simulation with lighter quarks, or whether the expected transition line reappears, for $m=695$ MeV, at temperatures below 129 MeV.  We reserve these questions for later study.

\section*{Acknowledgments}
The numerical simulations were performed at the Phoenix and Vienna
Scientific Cluster (VSC) at Vienna University of Technology. This research
was supported by the Erwin Schr\"odinger Fellowship program of the Austrian
Science Fund FWF (``Fonds zur F\"orderung der wissenschaftlichen Forschung'')
under Contract No. J3425-N27 (R.H.) and the U.S.\ Department of Energy under
Grant No.\ DE-FG03-92ER40711 (J.G.). R.H. is now supported by DFG under SFB/TRR55.

\bibliography{Lattice2017_163_HOELLWIESER}

\end{document}